\documentclass[a4paper,twocolumn,showpacs,superscriptaddress,floatfix]{revtex4-1}
\usepackage{latexsym}
\usepackage{amssymb}
\usepackage{amsfonts}
\usepackage{amsmath}
\usepackage{bm}
\usepackage{graphicx}
\usepackage{color}
\usepackage{subfigure}
\usepackage{times}
\usepackage{units}
\usepackage{hyperref}

\usepackage[utf8x]{inputenc}
\usepackage{amssymb,amsmath}
\usepackage[squaren]{SIunits}
\graphicspath{{./plots/}}

\newcommand{\Fig}[1]{Fig.~\ref{#1}}

\newcommand{\be}{\begin{equation}}
\newcommand{\ee}{\end{equation}}
\newcommand{\bea}{\begin{eqnarray}}
\newcommand{\eea}{\end{eqnarray}}
\newcommand{\beqn}{\begin{eqnarray*}}
\newcommand{\eeqn}{\end{eqnarray*}}
\newcommand{\ba}{\begin{align}}
\newcommand{\ea}{\end{align}}

\newcommand{\cf}{\textit{cf.}~}
\newcommand{\ie}{\textit{i.e.,}~}
\newcommand{\eg}{\textit{e.g.,}~}

\begin{document}


\title{On the black hole from merging binary neutron stars: how fast can it spin?}

\author{Wolfgang \surname{Kastaun}}
\affiliation{Max-Planck-Institut f\"ur Gravitationsphysik, Albert-Einstein-Institut, Potsdam, 14476, Germany}

\author{Filippo \surname{Galeazzi}}
\affiliation{Departamento de Astronom\'{\i}a y Astrof\'{\i}sica, Universitat de Val\`encia, Dr. Moliner 50, 46100, 
Burjassot (Val\`encia), Spain}
\affiliation{Max-Planck-Institut f\"ur Gravitationsphysik, Albert-Einstein-Institut, Potsdam, 14476, Germany}

\author{Daniela \surname{Alic}}
\affiliation{Max-Planck-Institut f\"ur Gravitationsphysik, Albert-Einstein-Institut, Potsdam, 14476, Germany}

\author{Luciano \surname{Rezzolla}}
\affiliation{Max-Planck-Institut f\"ur Gravitationsphysik, Albert-Einstein-Institut, Potsdam, 14476, Germany}

\author{Jos\'e A.~\surname{Font}}
\affiliation{Departamento de Astronom\'{\i}a y Astrof\'{\i}sica, Universitat de Val\`encia, Dr. Moliner 50, 46100, 
Burjassot (Val\`encia), Spain}


\begin{abstract}
The merger of two neutron stars will in general lead to the formation of
a torus surrounding a black hole whose rotational energy can be tapped to
potentially power a short gamma-ray burst. We have studied the merger of
equal-mass binaries with spins aligned with the orbital angular momentum
to determine the maximum spin the black hole can reach. Our initial data
consists of irrotational binaries to which we add various amounts of
rotation to increase the total angular momentum. Although the initial
data violates the constraint equations, the use of the constraint-damping
CCZ4 formulation yields evolutions with violations smaller than those
with irrotational initial data and standard formulations.  Interestingly,
we find that a limit of $J/M^2 \simeq 0.89$ exists for the dimensionless
spin and that any additional angular momentum given to the binary ends up
in the torus rather than in the black hole, thus providing another
nontrivial example supporting the cosmic censorship hypothesis.
\end{abstract}


\maketitle



\emph{Introduction.~}A long-standing open question in general relativity
is how close the angular momentum $J_{_\text{BH}}$ of a black hole (BH)
created in astrophysical scenarios can get to the theoretical maximum,
which for an isolated BH is $M_{_\text{ADM}}^2$ (throughout this article 
we use geometric units unless noted otherwise), with $M_{_\text{ADM}}$
the ADM mass. Under very general conditions the limit is $A/(8 \pi)$, 
with $A$ the horizon area~\cite{JaramilloReiris:2011}. So far, no 
violation of the cosmic censorship hypothesis, \ie no formation of a 
naked singularity, has been found in numerical simulations evolving 
generic initial data satisfying the dominant energy condition (but see
also~\cite{Bode2011b}). This question has been studied for the collapse of
supermassive stars, \eg \cite{Saijo2009}, neutron stars (NSs) with
$J_{_\text{NS}}/M_{_\text{NS}}^2>1$ \cite{Giacomazzo2011}, for the merger
of binary NSs (BNSs), \eg \cite{Baiotti08,Kiuchi2009,Gold2012}, of BH-NS
binaries, \eg \cite{Kyutoku2011c}, and of BH binaries,
\eg\cite{Hinder:2007qu}. In all cases investigated, the final BH spin is
below the critical value and semi-analytical estimates seem to indicate
this is true for all configurations~\cite{Barausse:2009uz,
  Pannarale2012}.

Determining the final spin produced in the merger of BNSs is not a mere
academic question as the rapid rotation of the BH is a key ingredient in
all of the models in which the BNS merger is thought to lead to a jet
formation and a short gamma-ray
burst~\cite{Nakar:2007yr,Lee:2007js,Rezzolla:2011}. Because the
energetics of the emission will depend sensitively on the BH spin, an
accurate measure of the maximum value attainable can help set upper
limits on the efficiency of the process (see~\cite{Giacomazzo2012b} for a
recent discussion and a list of references). Despite its importance, this
issue has not been addressed yet and the main reason is the lack of
constraint-satisfying initial data for spinning NSs. Early attempts to
construct such initial data, \eg~\cite{Marronetti03, Baumgarte:2009}, did
not have a satisfactory solution of the Euler equation according
to~\cite{Tichy11}. Only recently, initial data for BNSs with spins has
been computed~\cite{Tichy12}, but no evolutions have been reported yet.

Here we follow a novel approach, which consists of setting up
constraint-violating initial data by adding a rotational velocity field
to self-consistent irrotational initial data, and then evolving with a
recently developed constraint-damping CCZ4 formalism
\cite{Alic:2011a}. We show that this choice reduces the constraint
violations quickly to a level which is even smaller than the one
encountered in BNS evolutions of irrotational initial data with standard
formulations. The drawback of this simple method is that the artificial
spin-up also introduces oscillations of the star and eccentricity of the
orbit. This would be unacceptable for modelling gravitational-wave
emission, but it is adequate to assess the influence of the NS spin on
the BH.

\noindent\emph{Binary Initial Data with Spin.~} The best way to decide
what are realistic expectations for the spin in merging BNSs is to look
at the available observations. Although the distribution of NS spins is
still unknown, NSs at birth are expected to have spins in the range
$10$--$140$ ms~\cite{lorimer:2008}, and dimensionless spin parameters
$J_{_\text{NS}}/M_{_\text{NS}}^2\sim 1$ should be attainable only by
millisecond pulsars that have been spun-up by
accretion~\cite{Bildsten97}. It is however unlikely that the old NSs of a
binary system which is about to merge have periods less than 1 ms, and
all the observational evidence supports this conclusion. We know, in
fact, that the period of the fastest known pulsar in a BNS system,
J0737-3039A, is $22.70 \usk\milli\second$, yielding a dimensionless spin
parameter $J_{_\text{NS}}/M_{_\text{NS}}^2\sim 0.05$~\cite{Brown12}. 
In view of these considerations we restrict our analysis to binaries 
$-0.2 \lesssim J_{_\text{NS}}/M_{_\text{NS}}^2\lesssim 0.3$, thus well 
within the range of realistic possibilities.

To construct initial data for a binary system of spinning NSs, we modify
the velocity field of the irrotational equal mass solutions computed
using the \texttt{LORENE} code \cite{Gourgoulhon:2000nn}. We set
$\vec{\boldsymbol{w}} = \left(1-s\right) \vec{\boldsymbol{w}}_L +
s\vec{\boldsymbol{\Omega}} \times \vec{\boldsymbol{x}}$, where $w^i =
u^i/u^0$ denotes the coordinate velocity of the fluid in the star,
$u^{\mu}$ the fluid 4-velocity, $\vec{\boldsymbol{\Omega}}$ the orbital
angular velocity vector, $\vec{\boldsymbol{w}}_{_L}$ is the original
irrotational velocity field, and $s$ is a free parameter we tune to
increase/decrease the spin of the star $J_{_\text{NS}}$ ($s=0$
corresponds to a purely irrotational binary, $s=1$ to a corotating one).
Once a spin-up/down is introduced, we parametrize a binary in terms of
the additional ADM angular momentum it has with respect to the
irrotational model, $\Delta J_{_\text{ADM}}$. Note we compute
$J_{_\text{ADM}}$ using Eq.~(68) of~\cite{Gourgoulhon:2000nn} because
this form is more robust against constraint violations. Two remarks
should be made on our prescription for $\vec{\boldsymbol{w}}$. First, it
corresponds to stars with spins aligned with the orbital angular
momentum, which is the most relevant case for our goals, since it leads
to the maximum increase of total angular momentum. Second, it guarantees
that the residual vector field is orthogonal to the density gradient, and
hence the flow is adapted to the deformed shape of the stars in the
binary. However, because the hydrostatic equations are no longer
satisfied with the new velocity field, it introduces oscillations in the
stars as they are evolved. The relative variation of the central
rest-mass density stays below 1\% for the irrotational models, is $\le 8\%$
for $s=0.85$, and $\le 15\%$ for the extreme case of $s=1.2$.

Using the above method, we construct sequences of increasing spin, each
with a fixed baryon mass, from $1.625M_\odot$ to $1.901M_\odot$ for each
star, and the same initial separation of $45\usk\kilo\meter$, which
corresponds to roughly $3$ orbits until merger for the irrotational case
(up to $4$ for the fastest spinning cases). The stars obey initially a
polytropic equation of state (EOS) $P(\rho)=K\rho^\Gamma$ with $\Gamma=2$
and $K=123.6$ in units in which $G=c=M_\odot=1$. During the evolution,
however, we use an ideal-fluid ($\Gamma$-law) EOS~\cite{Baiotti08} with
$\Gamma=2$. Since we are interested in the \emph{maximum} spin of the BH,
we focus on binaries that will collapse promptly as any intervening
long-lived hypermassive neutron star (HMNS) would just extract additional
angular momentum from the system. Hence, we consider only models with
total baryon masses well above the maximum one for a nonrotating star,
which for our EOS is $2.0\,M_\odot$. We increase the spin frequency up to
55\% of the Kepler limit, which corresponds to $s=1.2$, but larger values
are possible in principle. As in our previous work
(\eg~\cite{Baiotti2011}), we use adaptive mesh-refinement
techniques~\cite{Schnetter-etal-03b} with 6 levels during inspiral, two
of which follow the stars, and a 7th finest level activated at the time of
collapse. The outer boundary is located at 756 km, where Sommerfeld
radiation boundary conditions are applied.

\begin{figure*}
  \centering
  \includegraphics[width=1.099\columnwidth]{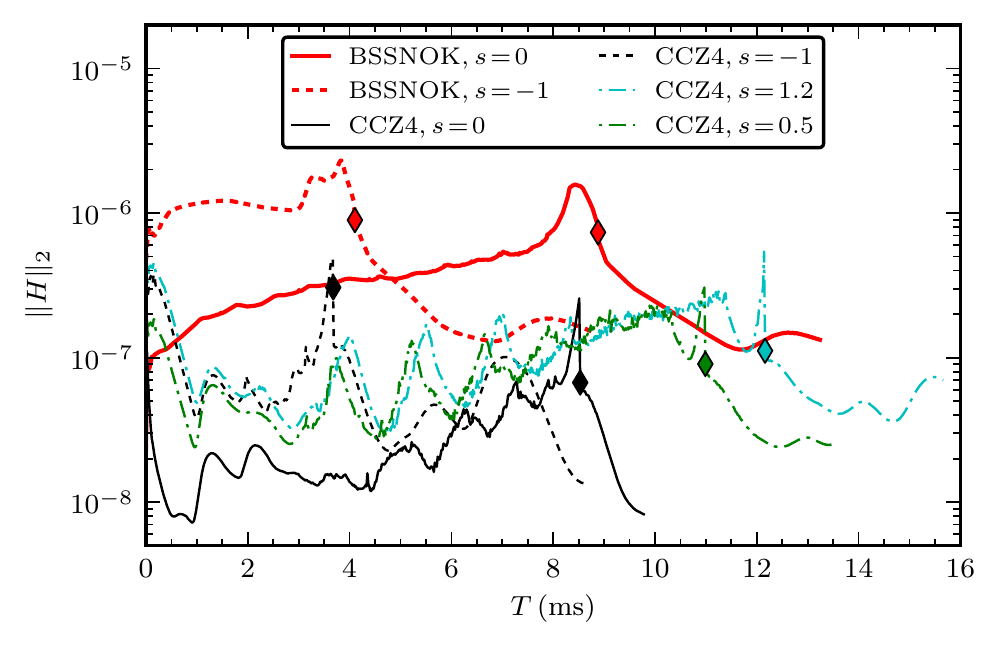}
  \includegraphics[width=0.89\columnwidth]{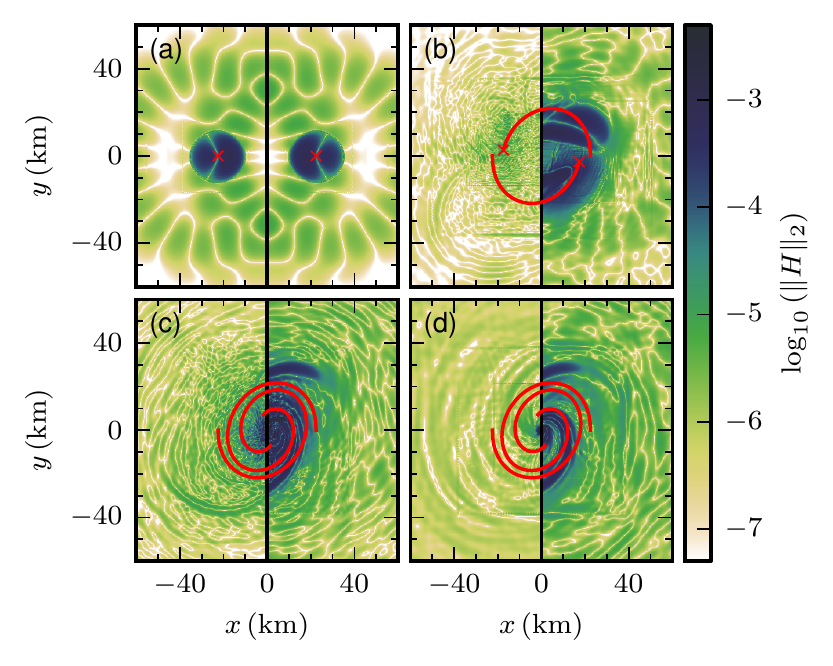}
  \caption{\emph{Left.~}Time evolution of the $L_2$-norm of the
    Hamiltonian constraint $H$ over the whole domain, excluding the AH,
    obtained using CCZ4 and BSSNOK formulations.  The symbols mark the AH
    formation.  \emph{Right.~} Snapshots in the orbital plane for $s=-1$,
    at times (a) $t=0$, (b) $t=1.48\usk\milli\second$, during inspiral,
    (c) $t=3.48\usk\milli\second$, shortly before horizon formation, (d)
    $t=4.46 \usk\milli\second$, shortly after.  The left half of each
    panel shows the $L_2$-norm of $H$ as obtained with CCZ4 evolutions,
    the right half with BSSNOK. The red solid lines mark the trajectories
    of the NSs. In all cases the binaries have $M_{_\text{NS}} = 1.78\usk
    M_\odot$.}
  \label{fig:constraints}
\end{figure*}

\noindent\emph{CCZ4 and BSSNOK formulations.~}We have already mentioned
the importance of using a formulation of the Einstein equations that
damps rapidly the violations of the constraints caused by our
modifications of the initial data. To achieve this, we employ the
recently proposed conformal and covariant Z4 formulation
(CCZ4)~\cite{Alic:2011a} and couple it to the equations of relativistic
hydrodynamics. We have carried out a systematic investigation of this new
formulation in comparison with the more standard BSSNOK
one~\cite{Baumgarte99, Alcubierre:2008}, including the evolution of a 
stable isolated NS, the collapse of an unstable star to a BH, and the 
merger of BNSs in quasi-circular and eccentric orbits. For lack of space 
these tests will be presented in detail in a forthcoming longer 
publication and here we restrict the discussion to the constraint-damping 
properties of CCZ4. The comparison reveals not only the advantages of 
CCZ4 over BSSNOK. It also provides a way to estimate how much constraint 
violations are tolerable before the influence on the dynamics becomes 
significant, by looking at the difference between two simulations of the 
same system with very different constraint violations.
More specifically, we can start assessing the role played by the initial
constraint violations with the extreme case of an eccentric binary as
created from quasicircular initial data in which we reduce artificially
the linear momenta by 15\%. Doing this induces very large constraint
violations in both formulations, although the $L_2$-norm of the
Hamiltonian constraint in CCZ4 evolutions always becomes smaller by a 
factor 5--10. Not surprisingly, the dynamics of the system shows large
differences in the phase evolution and in the trajectories, demonstrating
that the initial constraint violations are too high to yield meaningful
results. On the other hand, for quasicircular binaries, the orbital
trajectories for CCZ4 and BSSNOK agree quite well, with a phase error of
4\%. At the time of merger, the mass and the spin of the BH agree within
0.6\% and 0.4\%, which is less than our estimate of 1\% for the numerical
errors. Again, we find that the use of CCZ4 reduces the $L_2$-norm of the
Hamiltonian constraint, on average by one order of magnitude. In
particular, a very sharp decrease is seen during the first millisecond
(\cf left panel of \Fig{fig:constraints}). Under similar but not
identical conditions, Ref.~\cite{Hilditch2012} has reported a larger
decrease for an alternative conformal formulation of the Z4
system, Z4c. Without a direct comparison it is difficult to assess why 
it results in a smaller violation. It may also be due to the improved
treatment of the outer boundaries in~\cite{Hilditch2012}.

A similar behaviour is also observed for NSs that are spun-up/down. This 
is shown in the left part of \Fig{fig:constraints}, which reports the evolution of the
$L_2$-norm for a number of binaries with different degrees of
spin-up/down, when evolved with either the CCZ4 or the BSSNOK
formulation. The initial violation is obviously larger for larger changes
of spin, but in all CCZ4 evolutions the violation is reduced by about an
order of magnitude after 1 ms, becoming less than the one from BSSNOK
evolutions with constraint-satisfying initial data. Afterwards, the
constraints exhibit the usual behavior resulting from the interplay
between constraint-violating numerical errors and the action of the
damping terms in the CCZ4 system~\cite{Alic:2011a}, resulting in a
series of maxima and minima. This behaviour should be contrasted with the
one of the BSSNOK formulation, where the constraints instead remain large
during the whole evolution. For the latter, the errors tend to accumulate
at each point in space, causing the stars to leave behind lumps of
constraint violations. This is shown in the right part of
Fig.~\ref{fig:constraints}, which displays the Hamiltonian constraint
violation at different stages of the evolution for the case $s=-1$, both
for CCZ4 and BSSNOK. The comparison highlights that while the constraint
violations are damped/propagated away in the CCZ4 evolution, they remain
static in the BSSNOK one.

Interestingly, even with the most severe spin change of the $s=-1$
binary, the violation with the CCZ4 formulation is smaller by a factor
$\sim 5$ than the one obtained with the BSSNOK formulation for
irrotational binaries (red solid line), which represents the best-case
for that formulation. Therefore, since in all CCZ4 evolutions the
constraint violations fall below the irrotational case with BSSNOK, we
conclude that after the first millisecond the errors due to constraint
violations are comparable with the differences between BSSNOK and CCZ4 in
the irrotational case. Furthermore, to assess also the influence of the
large violation during the first millisecond we note that for the case
$s=-1$, the difference in BH mass, spin, and merger time between
BSSNOK and CCZ4 evolutions are 0.7\%, 3\%, and 10\%,
respectively. Since CCZ4 evolutions always have smaller violations
after the first millisecond, we conclude that the error in the BH mass
and spin due to constraint violations to be less than 1\%, and the total
error, including the numerical one, to be less than 2\%.

We note that the action of the constraint-damping CCZ4 formulation is
that of mapping the constraint-violating initial data closer to the
space of self-consistent solutions of the Einstein-Euler equations.
This mapping is active throughout the evolution and not only during the
first millisecond, although that is when it is most evident. We also
note that the use of constraint-violating initial data inevitably also
changes the physical parameters of the system, such as the eccentricity and
the hydrostatic equilibrium (which is perturbed). Although we lose a
certain amount of control over the physical parameters of the system
this way, part of the uncertainty also comes from the fact that the
added rotational velocity field affects also other physical quantities
besides the intended one, which is the spin. It is for those
reasons that we cannot prescribe the physical properties of our initial
data exactly. The uncertainty introduced is larger for systems with the
largest spins. This is why our simulations cannot be used for a
reliable description of gravitational waveforms and why we do not
consider stars spinning near the Kepler limit. However, because our
goal here is to show that a limit exists to the maximum spin of the
black hole, we are not particularly concerned whether our initial data
corresponds exactly to a particular configuration as long as it is
physically consistent.

\noindent\emph{Final BH spin.~}We turn now to discuss our physical
results and their astrophysical implications. We first note that
for the lightest binary ($M_{_\text{NS}}=1.63\usk M_\odot$), only the irrotational 
model forms a BH promptly, while a HMNS is formed for $s=0.5, 1.2$. 
This is not
unexpected, but highlights that if the total mass is close to the
critical mass for a prompt collapse, the spin of the NSs can have a
strong impact on the dynamics of the merger and thus on the
gravitational-wave signal. For the heavier binaries, instead, we observe
the immediate formation of a BH and a torus. Quasi-stationarity is
reached soon after merger, with the BH mass and spin changing less than
0.4\% during the last $50\,M_{_\text{BH}}$ for all simulations, and with
negligible mass accretion. These conditions are essential to use the
isolated-horizon formalism to measure the properties of the BH.

To measure the contribution to the total angular momentum of the fluid
orbiting outside the BH we monitor the quantity $J_{_\text{F}} \equiv
-\int_{V_o} d^3V n^{\mu} \phi^{\nu} T_{\mu\nu}$, where $V_o, d^3V,
n^{\mu}, T_{\mu\nu}$ denote, respectively, the portion of the time slice
$\Sigma_t$ outside the apparent horizon (AH), the proper volume element,
the unit normal to $\Sigma_t$, the energy-momentum tensor, and
$\boldsymbol{\phi} \equiv \boldsymbol{\partial}_\phi$ is the basis vector
of the spherical coordinate system obtained from the Cartesian one used
in the simulations. $J_{_\text{F}}$ reduces to the angular momentum in
the Newtonian limit, but is not conserved unless $\boldsymbol{\phi}$ is a
Killing vector (in which case it is conserved even for non-axisymmetric
flows~\cite{Kastaun2011}) and coincides with the Komar angular momentum
for stationary axisymmetric spacetimes. We define the angular momentum 
of the torus $J_{_\text{T}}$ as the angular momentum $J_{_\text{F}}$ of 
the fluid at the end of the simulation, when the BH has become stationary 
and mass accretion is negligible. 

On the other hand, to measure the angular momentum of the BH, we use the
isolated-horizon formalism \cite{Ashtekar01a,Dreyer02a}, computing the
integral $J_{_\text{BH}}(t) \equiv \left(8 \pi \right)^{-1} \int_{A_t}
\!\!d^2 V \Phi^{\mu} R^{\nu} K_{\mu\nu}$ on the AH surface $A_t$, where
$\Phi^{\mu}, R^{\nu}, K_{\mu\nu}$ are, respectively, an (approximate)
axial Killing vector on the AH, the unit normal to $A_t$ on $\Sigma_t$,
and the extrinsic curvature. Surprisingly, the sum of $J_{_\text{F}}$ and
$J_{_{\text{BH}}}$ at the end of the simulations agrees with
$J_{_\text{F}}$ at the time of AH formation to better than 2.2\% on
average, with a variance $\approx 0.5\,\%$ during the whole time from the
formation of the horizon until stationarity (the surprise comes from the
fact that $J_{_\text{F}}$, which is not expected to be conserved in
general, nevertheless represents a rather useful measure at the time the 
black hole is formed). This behaviour can be explained using Eq.~(45)
in~\cite{Hayward06}, which indicates that the time derivative of this sum
is small if $\phi^{\mu}$ becomes approximately a Killing vector of the
spacetime and the gravitational-wave emission from the AH is small. Both
conditions are satisfied at the time of horizon formation to a degree
sufficient for us to use $J_{_\text{F}}$ at this time to estimate the
final total angular momentum.

\begin{figure}
  \centering
  \includegraphics[width=0.98\columnwidth]{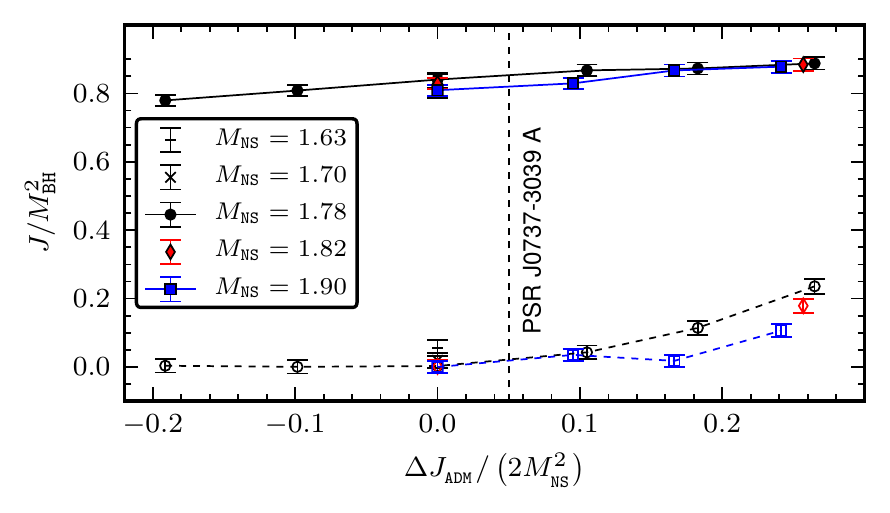}
  \caption{Dimensionless spin of the BH $J_{_\text{BH}} /
    M^2_{_\text{BH}}$ (filled symbols) and of the torus $J_{_\text{T}} /
    M^2_{_\text{BH}}$ (empty symbols) versus the additional initial
    angular momentum $\Delta J_{_\text{ADM}}$. The error-bars include 
    discretisation errors and the influence of the constraint 
    violations.}
    \label{fig:evol_angmom}
\end{figure}

The results of our simulations are summarized in the left panel of
Fig.~\ref{fig:evol_angmom}, which reports with filled symbols the
dimensionless spin of the BH, $J_{_\text{BH}} / M^2_{_\text{BH}}$.  Note that
$M_{_\text{BH}}$ is the horizon mass \cite{Dreyer02a} computed from
$J_{_\text{BH}}$ and the horizon area using the Kerr formula, so that the
extremal case $8 \pi J_{_\text{BH}} = A$ \cite{JaramilloReiris:2011} is
equivalent to $J_{_\text{BH}} = M^2_{_\text{BH}}$. The data is presented
as a function of the (dimensionless) additional initial angular momentum
per star, $\tfrac{1}{2}\Delta J_{_\text{ADM}}/M^2_{_\text{NS}}$, relative
to the corresponding irrotational model. Values of $\Delta
J_{_\text{ADM}}/M^2_{_\text{NS}} \gtrsim 0$ correspond to an increase of
total angular momentum, while $\Delta J_{_\text{ADM}}/M^2_{_\text{NS}}
\lesssim 0$ to NSs with antialigned spins. For comparison, we also plot
the estimate for the dimensionless spin of PSR J0737-3039A.

As it is natural to expect, the BH spin increases with the initial total
angular momentum of the binary, and for a binary like PSR J0737-3039A it
is not much larger than for an irrotational one.  Surprisingly, however,
the growth is not constant and actually saturates for $\Delta
J_{_\text{ADM}}/M^2_{_\text{NS}} \simeq 0.2$, quite independently of the
initial (baryon) mass in the binary. The largest BH spin we obtain is
$J_{_\text{BH}}/M^2_{_\text{BH}} = 0.888 \pm 0.018$ for
$M_{_\text{NS}}=1.78\,M_{\odot}$ and $s=1.2$. At the same time, we also
find that the combined angular momentum of the BH and of the torus,
$J_\text{tot} = J_{_\text{BH}} + J_{_\text{T}}$, continues to increase
for all values of $s$ used. More precisely, for $s \lesssim 1$, the
increase is essentially linear, with $J_\text{tot} =
J_\text{tot}^\text{irrot} + \eta \Delta J_{_\text{ADM}}$, where $\eta =
0.55\,(0.60)$ for $M_{_\text{NS}} = 1.78\,(1.90) M_\odot$ and
$J_\text{tot}^\text{irrot}$ is the value of $J_\text{tot}$ for the
irrotational binary. This means that roughly half of the additional
angular momentum has been radiated as gravitational waves
(see~\cite{Rezzolla:2010} for a detailed discussion of the angular
momentum radiated in gravitational waves for simulations similar to the
ones discussed here). For sufficiently large spin-ups, a significant
fraction of the remaining angular momentum is transferred to the torus,
which therefore acts as the channel absorbing the excess angular momentum
and limits the spin-up of the BH. This is reported in
Fig.~\ref{fig:evol_angmom}, which shows that $J_{_\text{T}}$ (empty
symbols) is very small for antialigned spins and irrotational binaries,
but increases rapidly with the initial stellar spin.

Of course it is difficult to provide a simple explanation for this
nonlinear behaviour, although one might speculate that spinning up the
NSs will provide matter with sufficient angular momentum to then produce
a torus. Overall, however, our results provide another nontrivial example
that, as proposed by the cosmic censorship hypothesis, a self-consistent
evolution of the Einstein equations from generic initial conditions leads
to a BH formation rather than a naked singularity.

We will conclude with a few remarks. First, it is reasonable that the
material in the torus will eventually be accreted onto the BH,
transferring angular momentum and further increasing the BH spin. 
However, this will happen on dissipative timescales which are longer and
thus not relevant for the central engine of gamma-ray bursts, which
should be ignited on a dynamical timescale after the merger. The tori 
produced in these simulations, in fact, are expected to be accreted on a
timescale of $\sim 0.3 \usk\second$~\cite{Rezzolla:2011} 
(\cite{Rezzolla:2010} indicates that the accretion timescale becomes
even larger for the case of unequal masses). Second, because magnetic
fields are expected to further decrease the angular momentum in the
system they are not particularly relevant in our considerations, which
focus on the maximum spin in a prompt BH formation. Finally, although
the BH spin seems to increase for irrotational 
unequal-masses~\cite{Rezzolla:2010}, the spin could also decrease 
depending on the amount of mass and spin in the torus. In any case, we
expect an upper limit will be found also for generic binaries and this 
will be the focus of our future work.


\begin{acknowledgments}
Support comes through the DFG grant SFB/Trans-regio 7, ``CompStar'', a
Research Networking Programme of the ESF and the Spanish MICINN (AYA
2010-21097-C03-01). FG is supported by a VESF fellowship from the
European Gravitational Observatory (EGO-DIR-69-2010).
\end{acknowledgments}

\bibliographystyle{apsrev4-1-noeprint}
\bibliography{aeireferences.bib}

\end{document}